\documentclass[a4paper,10pt]{article}
\usepackage[english]{babel}
\usepackage[utf8]{inputenc}
\usepackage[]{hyperref}
\usepackage[pdftex]{graphicx}
\usepackage{listings} 
\usepackage{fancyvrb}
\usepackage{listings}
\usepackage[usenames,dvipsnames]{color}

\lstdefinelanguage{JavaScript}{
  classoffset=0,
  sensitive=true,
  morekeywords={attributes, class, classend, do, empty, endif,
    endwhile, fail, function, functionend, if, implements, in,
    inherit, inout, not, of, operations, out, return, set, then,
    types, while, use,protocol, package, static,slots,extends,mixins}, 
  keywordstyle=\color{NavyBlue}\bfseries,
  classoffset=1,
  sensitive=true,
  morekeywords={prototype,this},
  keywordstyle=\color{ForestGreen}\bfseries,
  classoffset=2,
  sensitive=true,
  morekeywords={Class},
  keywordstyle=\color{OrangeRed}\bfseries,
  stringstyle=\color{red}\ttfamily
}

\title{JSC : A JavaScript Object System}

\author{Artur Ventura\\\url{artur.ventura@ist.utl.pt}}

\begin{document}
\maketitle

\begin{abstract}
The JSC language is a superset of JavaScript designed to ease the development of large web applications. This language extends JavaScript's own object system by isolating code in a class declaration, simplifying multiple inheritance and using method implementation agreements.  
\end{abstract}

\section{Motivation}

As web applications have been gaining more dynamic behavior, JavaScript has become more
important in web development. As such, it is useful to use a software
engineering aproach to JavaScript.\\

JavaScript object-oriented model is prototype-based. This model can be
extremely versatile to develop. However, in a large team of
developers, not only problems such as name colisions and unintentional method
redefinition can occur, but also maintenance becomes an issue.\\

JSC attempts to remedy some of these issues by implementing a Object
System on top of standard JavaScript.
\section{Related Work}
There have been some attemps to add structed programming to
Javascript.\\

In ECMAScript (language from witch JavaScript is a dialect) last
version were added support for defining classes. Another dialect of
ECMAScript, ActionScript supports defining classes. Althought those
systems are more advanced (they support typechecking via anotation on
variables, generics, dynamic classes, templates, etc.), they aren't
avaible in the current browsers. It's expected that future versions of
JavaScript extends this behavior from ECMAScript.\\

Other attempt have been creating creating supersets to JavaScript. One
example of this is Objective-J. This is a sctrict superspet of
Javascript, but adds a object system by implementing a language
similar to Objective-C. This aproach is similar to JSC, but this
language uses the same message passing than Objective-C making the
methods unparsable by JavaScript.

\section{Object Oriented JavaScript}

In JavaScript, a class is a function, so its definition is similar to
the definition of a function, as shown next:
\\
\begin{lstlisting}
function Rectangle(w, h) {
    this.width = w;
    this.height = h;
}
\end{lstlisting}

A function in javascript is of type \texttt{Function} but behaves similarly to a
hash or a dictionary. One of the properties in this object is the
\texttt{prototype}. This property represents a structure that is not
yet fully defined and it is only formed when it is
instantiated. Instantiation can be done by:  
\\
\begin{lstlisting}[numbers=none]
new Rectangle(10,10);
\end{lstlisting}

\texttt{new} operator will invoke the function \texttt{Rectangle} with
\texttt{this} as a new object with a reference to
\texttt{Rectangle.prototype}, and return this object\footnote{Actually
\texttt{Rectangle} could return anything, but the default is
\texttt{this}}.Any changes made to this object will only afect the new
object, and not the prototype. If a property is not defined in the object but on
the prototype, then the prototype's value is returned.\\

A method can be defined by adding a function to the class prototype as
shown next:
\\
\begin{lstlisting}
Rectangle.prototype.getArea = function (){
    return this.height * this.width;
}
\end{lstlisting}

Methods are called like this:

\begin{lstlisting}[numbers=none]
rectangle.getHeight()
\end{lstlisting}

As with the constructor, \texttt{getHeight} will be invoked with \texttt{this} as a
reference to the instance, in this case to \texttt{rectangle}. This
method will be avaible to all instances that don't define localy 
the method \texttt{getArea}.\\

The code below shows how inheritance is implemented: by copying the
\texttt{prototype} of the superclass to the subclass.
\\
\begin{lstlisting}
function PositionedRectangle(x, y, w, h) {
    Rectangle.call(this, w, h);

    this.x = x;
    this.y = y;
}
PositionedRectangle.prototype = new Rectangle();
\end{lstlisting}

\subsection{Problems}

The first problem is the inexistence of seperation between functional
and object oriented programming. All the examples shown are
extensions to JavaScript's functional programming that allows to
create a behavior similar to object oriented.\\

Another problem is that today browsers share the same enviroment will
all JavaScript files. It's possible to have the same method defined in
two distinct files. This can difficult debugging and maintenance.

\section{Objectives}
JSC is an extension to JavaScript and was developed with the
following goals in mind: 

\begin{itemize}
\item Create a new object-oriented language, but keep (as much as
  possible) JavaScript's syntax and semantics.
\item Provide mechanisms to maintain the code isolated in packages.  
\item Provide a meta-object protocol that allows an easy use of reflection
  and intersection, while keeping the code organized.  
\item Provide simple and easy to use multiple inheritance.
\item Provide an implementation agreement between classes, which is
  similar to Java's interfaces or to Objective-C's protocols. 
\end{itemize}

\section{JSC}
\subsection{Class Definition}
JSC is defined in classes. Each class is defined in a single file. A
simple JSC Class is show next:

\begin{lstlisting}
package UI.Component;

class Rectangle{
  slots:[height,width],
  Rectangle: function (w,h){
    this.setHeight(h);
    this.setWidth(w);
  },
  getArea: function (){
    return this.getHeight() * this.getWidth();
  }
}
\end{lstlisting}

Each class begins with a package declaration. A class is always
referred to by its package and class name (in this case
\texttt{UI.Component.Rectangle}). The body of the class is declared as
a JavaScript \texttt{Object}. The fourth line represents the slot
declaration. In JSC, class slots can't be directly used. To access a
slot, a getter and setter method is provided for each slot. The fifth and
ninth lines declare a constructor and a method, respectively.
A constructor is a method with the same name of the class. A
constructor is not required, but there can be only one by class.\\

It is important to notice that only the class header and the
slot definition aren't capable of being parsed by JavaScript.
\subsection{Instantiation}
JSC attempts to minimize the usage of the global enviroment. Actually, the only
global definition required is the function \texttt{Class}. This
function expects a string with the canonical class location (package and
name), and returns the meta-object representing such class. In our
example, \texttt{Rectangle}'s meta-class can be obtained by:
\\
\begin{lstlisting}[numbers=none]
Class("UI.Component.Rectangle")
\end{lstlisting}

In runtime, each meta-class contains, among others, a method called
\texttt{create}. This method will create a new instance and call the
class constructor. With this, we can now instantiate our
\texttt{Rectangle} class:
\\
\begin{lstlisting}[numbers=none]
Class("UI.Component.Rectangle").create(10,10)
\end{lstlisting}

\subsection{Inheritance}
Inheritance can be used by extending the superclass as shown next: 

\begin{lstlisting}
package UI.Component;

class PositionedRectangle extends UI.Component.Rectangle{
  slots:[x,y],
  PositionedRectangle: function (x,y,w,h){
    Class("UI.Component.Rectangle").init(this,w,h);

    this.setX(x);
    this.setY(y);
  }
}
\end{lstlisting}

Inheritance in JSC works like \textit{mixins}. Each class method and
slot from the superclass will be added to the subclass. A class can
extend from several classes. In \texttt{PositionedRectangle} both the
height and with slot, getters and setters from \texttt{Rectangle} were
copied. The token \texttt{extends} insted of \texttt{mixin}  was used
because makes the class header similar to ECMAScript V4 and Java's.\\

Each meta-class object contains a \texttt{init} method that calls the
constructor for that class, using the first argument as the
instance. This is similar how Python's superclass constructor is
called.

\subsection{Static Enviroment}
Each JSC class can declare methods that can be used on a meta-class level. A
simple example is shown next:
\\
\begin{lstlisting}
package Main;

class App {
  static:{
    main: function (args){
      ...
    }
  }
}
\end{lstlisting}
\subsection{Protocols}

A protocol in JSC assures the existence of certain methods. A
protocol declaration is shown next:
\\
\begin{lstlisting}
package UI.Component;

protocol Draggable {
  element: true,
  eventListener: false
}
\end{lstlisting}

In our example \texttt{Draggable} declares the existence of two
methods, \texttt{element} and \texttt{eventListener}. The keyword next
to the method name declares the need to implement it. \texttt{true}
declares that this method is required to be implemented in each class
that extends this protocol while \texttt{false} guarantees that if such
method is not implemented, an empty function with such name will be
provided.\\

 A protocol can be extended only from other protocols, and
a class can implement any number of protocols. The verification of
required methods is done both in run- and compile-time.

\subsection{Class initialization}

When JSC starts\footnote{See Usage for more details}, all classes in
the classpool are going to be initialized. This is done by invoking the
meta-class method \texttt{classInit}.\\

This method will compute the effective set of methods and slots that
this class possesses, and it will detect if all the protocols have
been satisfied. Finally, it will setup the prototype.\\ 

\subsection{Runtime intersection}
It is possible to create a new class/protocol, or change a
class/protocol by altering the meta-class definition, in any moment
during execution. However, it is required to call again
\texttt{classInit} so that the prototype is reconstructed.\\

It's also possible to extend the behavior of JSC by extending the
\texttt{lang.Class} class. Protocols is one example of this, but it's
quite simple to implement something similar to java's abstract classes
in JSC.

\section{Current Problems}
\subsection{Global Variables}
JSC attempts to minimize the usage of the global enviroment by encapsulating
the code into classes. However, However, JavaScript allows the
declaration of gobal variables from within a function.. Any variable
that is declared 
without the keyword \texttt{var} is declared in the global enviroment.
For instance:
\\
\begin{lstlisting}
function Foo(){
  var local = 1;
  global = 1;
}
\end{lstlisting}

Currently, this is not allowed in JSC, but it is not verified.

\subsection{Slot default value}
Currently, the slot declaration doesn't allow a slot to have a default
value. It would be nice to have a slot declaration similar to this:

\begin{lstlisting}
package Bar;

class Foo{
  slots:{
    aSlot:{ getter:"getSlot", setter:"setIt", default:1 },
    anotherSlot: { default: Class("Baz.Bing").create(1,2) }
  }
}
\end{lstlisting}

Currently, JSC doesn't support this syntax because there isn't a
easy to use parser of JavaScript available. The JSC compiler assures
the correctness of the code by loading the methods into a JavaScript Engine and
detecting the parsing errors generated.\footnote{\texttt{jscc} (JSC Compiler)
  currently uses Google's V8 Engine} \\

When such parser is available, both this and the Global Variables problem
can be addressed.

\section{Usage}
The current compiler can target 3 usage methods:
\begin{itemize}
\item \textbf{Server} - JSC was developed due to a need for
  developing a lot of code to run in a web server. In this mode, the code
  is loaded into a RDBMS and loaded upon need. 
\item \textbf{Client} - Another usage is generating a single classpool
  image in a self contained JavaScript file and load it in a
  browser like an ordinary JavaScript file. The JSC code can be accessed
  from other JavaScript files by using \texttt{Class} function.
\item \textbf{JSC Virtual Machine} - \texttt{jscvm} is a small virtual
  machine implementation of JSC.
\end{itemize}

\section{Performance}
A small library, implemented in JSC and with 19 files, occupies 196
KB. After compilation targeting the browser ended with a 
single file occupying 56KB. This file takes from below a second in
Chrome and up to 2 seconds in Firefox.\\

After using the YUI Compressor\footnote{YUI Compressor is available at
  \url{http://developer.yahoo.com/yui/compressor/}. This application
  not only removes whitespaces, but also reduces the variable names.}
this file shrinked to 40Kb. It almost loaded instantaneously
both in Chrome and Firefox.\\
\end{document}